\documentstyle[12pt]{article}

\makeatletter
\def\eqnumsection{\@addtoreset{equation}{section}\def\theequation
{\arabic{section}.\arabic{equation}}}
\makeatother

\title{
\begin{flushright}
{\large Yaroslavl State University \\
        Preprint YARU-HE-97/03 \\
        hep-ph/9702326} \\[10mm]
\end{flushright}
       {\LARGE\bf Axion decay $a \to f \tilde f$ in a strong} \\ 
       {\LARGE\bf magnetic field}}

\author{{\Large\bf N.V.~Mikheev } \\[2mm]
        {\large\it
             Division of Theoretical Physics, Department of Physics,} \\
        {\large\it
             Yaroslavl State University, Yaroslavl 150000, Russia} \\
        {\large\it E-mail: mikheev@yars.free.net} \\[4mm]
        {\Large\bf and} \\[4mm]
        {\Large\bf L.A.~Vassilevskaya} \\[2mm]
        {\large\it
             Moscow Lomonosov University, V-952, Moscow 117234, Russia} \\
        {\large\it E-mail: vasilevs@vitep5.itep.ru}}

\date{}

\begin{document}

\maketitle

\begin{abstract}
The axion decay into charged fermion-antifermion pair 
$a ~\to f ~\tilde f$ is studied in external crossed and 
magnetic fields.
The result we have obtained can be of use  to re-examine 
the lower limit on the axion mass in case of possible existence 
of strong magnetic fields at the early Universe stage.
\end{abstract}

\vspace*{40mm}

\centerline{\large will be published in {\it Phys. Lett. B}}

\thispagestyle{empty}

\newpage

Peccei-Quinn ($PQ$) symmetry~\cite{P1} continues to be the most
attractive solution to the problem of CP-conservation in strong 
interactions. Therefore the axion~\cite{WW}, 
the pseudo-Goldstone boson associated with $PQ$-symmetry,
is the subject of constant theoretical and experimental investigations 
(see, for example,  ~\cite{Raf1,Raf2}).
The allowed range for the axion mass, and therefore for
the $PQ$-symmetry breaking scale, $f_a$:

\begin{equation}
m_a \simeq 6.3 \left[\frac{10^6 {\rm GeV}}{f_a}\right] {\rm eV},
\end{equation}

\noindent is strongly constrained by astrophysical and cosmological 
considerations which leave
a rather narrow window~\cite{Raf1,Raf2,Tur,P2}:

\begin{equation}
10^{-6} {\rm eV} \leq m_a \leq 10^{-3} {\rm eV}.
\label{eq:MA}
\end{equation}

\noindent The upper bound on $m_a$ arises from astrophysical 
arguments because axion emission removes energy from stars 
altering their evolution. If the axion has a mass near 
the lower limit, then axions play a very important role in the 
Universe and could be the Universe's cold dark matter.
Despite constant theoretical and experimental 
interest to the axion all attempts of its experimental 
detection have not a success yet. It can be naturally
explained by the fact that axions are very light, very weakly 
interacting (coupling constant $\sim f^{-1}_a$) and very long living.
The axion lifetime in vacuum is ~\cite{Raf1}:

\begin{equation}
\tau (a \to 2\gamma) \sim 6,3 \cdot 10^{48} \,s \,
\left ( {10^{-3} eV \over m_a } \right )^6 \; 
\left ( {E_a \over 1 MeV } \right ) .
\label{eq:T0}
\end{equation}

On the other hand, the investigation of the axion physics
in strong external electromagnetic fields could appear to be 
very effective in getting new limits on the axion parameters.
An external field can not only influence substantially 
on decay processes (for example, a strong catalyzing
effect of the external field on neutrino radiative decay
$\nu \to \nu' \gamma$~\cite{L1}, double radiative axion decay 
$a \to 2\gamma$ ~\cite{L2}) but open new channels forbidden 
in vacuum as well (the well known photon splitting into 
electron--positron pair $\gamma \to e^+ e^-$ ~\cite{Klep}
and into two photons $\gamma \to \gamma \gamma$ ~\cite{Adl}).

Notice that investigations in strong magnetic fields 
($B > B_e$, $B_e = m^{2}_{e}/e \simeq 0.44 \cdot 10^{14} \, G$
is the critical, so called, Schwinger value) are, in our opinion, 
of principle interest in astrophysics 
(the field strengths inside the astrophysical objects
could be as high as $10^{15} - 10^{18} G$
~\cite{Rud,Lip,Boc}) and cosmology of the early Universe
(a generation of primordial strong magnetic fields is possible 
through thermal fluctuation with magnetic field strengths 
of order of $10^{12} - 10^{15} G$ ~\cite{Lem} 
or $10^{13} - 10^{17} G$~\cite{Taj}).

In this work we study the field--induced axion decay into charged
fermion--antifermion pair $a \to f \tilde f$ in the magnetic field.
We note that this process is forbidden in vacuum 
when $m_a < 2 m_f$. However the kinematics of a charged 
particle in a magnetic field is that which allows to have 
both time--like and space--like total momentum for 
the charged fermion--antifermion pair.
For this reason this process becomes possible in the magnetic
field even for massless pseudoscalar particle (arion ~\cite{AU}, 
for example), as it occurs in 
photon splitting $\gamma \to e^+ e^-$ ~\cite{Klep}.

An amplitude of this decay can be immediately 
obtained from the Lagrangian of axion-fermion interaction:

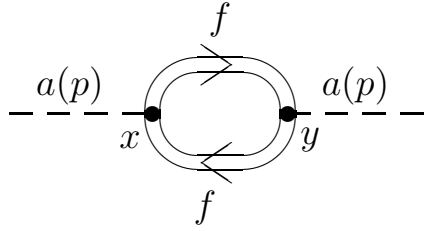
\begin{equation}
{\cal L}_{af} =  \frac{ g_{af}}{2 m_f} 
(\bar f \gamma_\mu \gamma_5 f) \, \partial_\mu \, a,
\label{eq:L0}
\end{equation}

\noindent using the known solutions of the Dirac equation in a 
magnetic field. Here $g_{af} = C_f m_f/f_a$ is a dimensionless 
coupling constant, $C_f$ is a model--dependent factor,
$m_f$ is a fermion mass. To obtain the decay probability one 
has to carry out a non-trivial integration over the phase 
space of the fermion pair taking the specific kinematics of 
charged particles in the magnetic field into account.

However the decay probability can be obtained by another easier way. 
It is known, that the decay probability of  
$a \to f \tilde f$ is connected with the imaginary part of the 
amplitude of the $a \to f \tilde f \to a$ transition via the 
fermion loop by the unitarity relation:

\begin{equation}
E_a W(a \to f \tilde f) = Im \,{\cal M}_{a \to a}.
\label{eq:W0}
\end{equation}

In the lowest order of the perturbation theory the amplitude 
${\cal M}_{a \to a}$ is described by the diagram
in the Fig.1 where double lines imply that the influence of the 
external magnetic field in the fermion propagator is taken into
account exactly. 

\begin{figure}[tb]

%%%%%%%%%%%%%%%%%%%% Fig.1 %%%%%%%%%%%%%%%%%%%%%%%%%%%%%%

\def\photonatomright{\begin{picture}(3,1.5)(0,0)
                                \put(0,-0.75){\tencircw \symbol{2}}
                                \put(1.5,-0.75){\tencircw \symbol{1}}
                                \put(1.5,0.75){\tencircw \symbol{3}}
                                \put(3,0.75){\tencircw \symbol{0}}
                      \end{picture}
                     }
\def\photonright{\begin{picture}(30,1.5)(0,0)
                     \multiput(0,0)(3,0){10}{\photonatomright}
                  \end{picture}
                 }
\def\photonrighthalf{\begin{picture}(30,1.5)(0,0)
                         \multiput(0,0)(3,0){5}{\photonatomright}
                      \end{picture}
                     }

%%%%%%%%%%%%%%%%%%%% end Fig.1 %%%%%%%%%%%%%%%%%%%%%%%%%%%%%%%%%%%%%%

\unitlength=1.00mm
\special{em:linewidth 0.4pt}
\linethickness{0.4pt}

\begin{picture}(60.00,45.00)(-40,10)
\put(35.00,32.50){\oval(20.00,15.00)[]}
\put(35.00,32.50){\oval(16.00,11.00)[]}

\put(44.00,32.50){\circle*{2.00}}
\multiput(44.00,32.50)(5.00,0.00){4}{\line(1,0){3.00}}

\put(26.00,32.50){\circle*{2.00}}
%\put(10.00,32.50){\photonrighthalf}
\multiput(7.00,32.50)(5.00,0.00){4}{\line(1,0){3.00}}

\put(36.50,39.00){\line(-3,2){4.01}}
\put(36.50,39.00){\line(-3,-2){4.01}}

\put(32.50,26.00){\line(3,2){4.01}}
\put(32.50,26.00){\line(3,-2){4.01}}

\put(15.00,35.00){\makebox(0,0)[cb]{\large $a (p)$}}

\put(35.00,45.00){\makebox(0,0)[cc]{\large $f$}}
\put(33.00,20.00){\makebox(0,0)[cc]{\large $f$}}

\put(53.00,35.00){\makebox(0,0)[cb]{\large $a (p)$}}

\put(23.00,29.00){\makebox(0,0)[cc]{\large $x$}}
\put(47.00,29.00){\makebox(0,0)[cc]{\large $y$}}

\end{picture}
\end{figure}

\noindent We stress that in order to obtain the correct expression
for ${\cal M}_{a \to a}$ we have to use a derivative nature of the 
axion-fermion interaction Lagrangian ~(\ref{eq:L0}) as it was 
first emphasized by Raffelt and Seckel ~\cite{Raf3}.
In general the expression for ${\cal M}_{a \to a}$ can be 
presented in the form:

\begin{equation}
{\cal M}_{a \to a} = - \,i \,\frac{g^2_{af}}{4 m_f^2} \,
\int d ^4 Z \, Sp[ S(-Z) \,\hat q \, \gamma_5 \, S(Z) \, 
\hat q \,\gamma_5)] \, e^{-i q Z},
\label{eq:M1}
\end{equation}

\noindent where  $S(Z)$ $(Z = x - y)$ is the translationally 
non-invariant part of the fermion propagator in the 
magnetic field ~\cite{L1}. The result of our calculations 
for the field--induced part of ${\cal M}_{a \to a}$ is:

\begin{eqnarray}
\Delta {\cal M}_{a \to a} & \simeq &  \frac{g^2_{af}}{8 \pi^2} 
\cdot \beta \, q^2_\perp \,
\int\limits_0^1 d u \, \int\limits_0^\infty d t \, 
e^{-i \Phi(u,t)} \,
{\cos \beta t - \cos \beta ut \over \sin \beta t},
\label{eq:M2}\\
\Phi(u,t) & = & 
\left (
m_f^2 - q^2_\perp \,{1 - u^2 \over 4} 
\right ) t 
+ {q^2_\perp \over 2 \beta} \,
{\cos \beta ut - \cos \beta t \over \sin \beta t},
\nonumber\\
q^2_\perp & = & \frac{e_f^2}{\beta^2} (q F F q) 
\simeq E_a^2 \sin^2 \theta,
\nonumber
\end{eqnarray}

\noindent where $\beta = \vert e_f B \vert = \sqrt {e_f^2(FF)/2}$;
$e_f = e Q_f$, $e > 0$  is the elementary charge,
$Q_f$ is a relative fermion charge; $F_{\alpha \beta}$ is the
external magnetic field tensor
($(q F F q) = q_\mu F^{\mu \nu} F_{\nu \rho} q^\rho $, 
$( F F ) =  F_{\mu \nu} F^{\nu \mu} $); 
$E_a$ is the energy of the decaying axion;
$\theta$ is the angle between the vectors 
of the magnetic field strength ${\vec B}$ and the momentum of
the axion ${\vec q}$. 

The general expression for the probability takes a rather 
complicated form. We present here the results of our 
calculations in two limiting cases which have a clear 
physical meaning: $q^2_\perp \gg \beta $ 
and $\beta > q^2_\perp$.

In the case $q^2_\perp \gg \beta $, when fermion and antifermion
with energies $E_f \gg \sqrt \beta$ can be born in the states
corresponding to the highest Landau levels,
the decay probability is:

\begin{eqnarray}
W  \simeq  \frac{ 3^{5/3}}  
{ 16 \pi^3 } \,\Gamma^4 \left (\frac{2}{3} \right ) \,
\cdot g_{af}^2 \,\frac{[e^2_f (q F F q)]^{1/3}}{E_a} \, \rho,
\label{eq:W1}
\end{eqnarray}

\noindent Here the factor $\rho$ is introduced which takes 
into account a possible influence of the medium on the axion decay 
($\rho = 1$ when the medium is absent):

\begin{eqnarray}
\rho  =  \frac {\int d W \, ( 1 - n_f ) \, ( 1 - n_{\tilde f} )}
{\int d W} ,
\label{eq:Rho}
\end{eqnarray}

\noindent where 

\begin{eqnarray}
n_f = \left [ exp \left ( \frac{E_f - \mu}{T} \right ) + 1
\right ]^{-1}
\nonumber
\end{eqnarray}

\noindent is the Fermi--Dirac distribution function of fermion at
a temperature $T$, $\mu$ is the chemical potential,
$n_{\tilde f}$ is the distribution function of antifermions.

\noindent The corresponding expression for the axion lifetime is:

\begin{eqnarray}
\tau \simeq \frac {2 \cdot 10^6}{\rho} 
\left ( \frac{ 10^{-13} }{ g_{af} } \right )^2
 \left ( \frac{ E_a }{ 100 MeV } \right )^{1/3} 
\left ( \frac{ 10^{15} G }{ |Q_f|\; B \sin{\theta} } 
\right )^{2/3} \; s.
\label{eq:T1}
\end{eqnarray}

\noindent As is seen from Eq. ~(\ref{eq:T1}) the strong 
magnetic field not only opens the decay channel but catalyzes 
substantially this process as well.

In the limiting case, $\beta > q^2_\perp$, the charged 
fermion--antifermion pair can be born only in the states 
corresponding to the lowest Landau level. This becomes possible 
if the high energy axion propagates almost along the magnetic 
field strength vector $\sin^2\theta \sim 4 m^2_f/ E^2_a$, 
so $E^2_a \sin^2\theta \sim 4 m^2_f < \beta $.
The decay probability has the form:

\begin{equation}
W \simeq \frac{g^2_{af}}{4 \pi} \, 
\frac{\beta}{E_a} \,
\frac{
\exp \left ( {\mbox{\large $
-\frac{ q^2_\perp}{2 \beta}$}}
\right )}
{\left({\mbox{\large $ 1 - 
\frac{4 m^2_f}{q^2_\perp }$}} \right)^{1/2}} \; \rho
\label{eq:W2}
\end{equation}

\noindent The expression for the lifetime is:

\begin{equation}
\tau \simeq \frac{1,4 \cdot 10^7}{\rho} \, 
\frac{1}{\vert Q_f \vert} \,
\left ( {10^{-13} \over g_{af}} \right )^2 \;
\left ( {10^{15} G\over B} \right ) \;
\left ( {E_a \over 100 MeV } \right ) \; 
\left( 1 - \frac{4 m^2_f}{q^2_\perp} \right)^{1/2} \; s.
\label{eq:T2}
\end{equation}

\noindent As it follows from Eq.~(\ref{eq:T2}), the axion 
lifetime tends to zero under the kinematical condition:
$q^2_\perp \to 4 m^2_f$. Actually the expression
$\left( 1 - 4 m^2_f /q^2_\perp  \right)^{1/2}$
has a minimal nonzero value due to the axion dispersion
in the magnetic field:

\begin{equation}
\left(
1 - \frac{4 m^2_f}{q^2_\perp}
\right)^{1/2}_{min}
\simeq \frac{1}{\sqrt{3}} \,
\left ( \frac{g^2_{af}}{2 \pi} \frac{B}{B_f} \right )^{1/3}.
\label{eq:SIN}
\end{equation}

\noindent So, the minimal value for the axion lifetime is:

\begin{equation}
(\tau)_{min} \simeq \frac{1,7 \cdot 10^{-2}}{\rho} \, 
\left ( {1 MeV \over \vert Q_f \vert m_f} \right )^{2/3} \;
\left ( {10^{-13} \over g_{af}} \right )^{4/3} \;
\left ( {10^{15} G\over B} \right )^{2/3} \;
\left ( {E_a \over 100 MeV } \right ) \; s .
\label{eq:TM}
\end{equation}

Considering possible applications of the result we have 
obtained to cosmology it is necessary to take an influence 
of a hot plasma into account. It is known that such an influence 
is reduced to suppressing statistical factors in the integral 
over the phase space of the fermion pair.
Under the early Universe conditions the hot plasma is 
nondegenerated one ($\mu \ll T$) and the medium parameter 
$\rho$ is inside the interval $\frac{1}{4} < \rho < 1$
for the expressions ~(\ref{eq:W1}) and ~(\ref{eq:T1}).
For fermions on the ground Landau level 
($E_f \sim E_a \sin \theta \ll T$) the parameter $\rho$
in ~(\ref{eq:W2}),  ~(\ref{eq:T2})
and  ~(\ref{eq:TM}).
is close to $\frac{1}{4}$. 

The result we have obtained could be of use to re-examine
the lower limit of the axion mass in the case of a possible 
existence of strong magnetic fields ($B > B_e$) at the early 
Universe stage.

\vspace*{5mm}

\noindent {\bf Acknowledgements}  

\vspace*{3mm}

The authors thank V.A.~Rubakov for fruitful discussions and 
A.V.~Kuznetsov for useful critical remarks.

\end{document}